\documentclass[usenatbib]{mn2e}
\usepackage{graphicx}
\usepackage{amsmath}

\title[Sub-structures in tidal tails]{Formation and Evolution of
  sub-structures in tidal tails: Spherical dark matter haloes} 

\author[Reinoso et al.]{
  B. Reinoso$^{1}$\thanks{E-mail: breinoso@udec.cl},
  M. Fellhauer$^{1}$\thanks{E-mail: mfellhauer@astro-udec.cl},
  R. V\'{e}jar$^{1}$\\
  $^{1}$Departamento de Astronom\'{i}a, Universidad de Concepci\'{o}n,
  Casilla 160-C, Concepci\'{o}n, Chile} 

\begin{document}

\label{firstpage}

\maketitle

\pagerange{\pageref{firstpage}--\pageref{lastpage}} \pubyear{2017}

\begin{abstract}
  Recently a theory about the formation of over-densities of stars
  along tidal tails of globular clusters has been presented, this
  theory predicts the position and time of formation of such
  over-densities and was successfully tested with N-body simulations of
  globular clusters in a point mass galactic potential.  In this work
  we present a comparison between this theory and our simulations
  using a dwarf galaxy orbiting two differently shaped dark matter
  halos to study the effects of a cored and a cuspy halo on the
  formation and evolution of tidal tails.  We find no difference using
  a cuspy or a cored halo, however, we find an intriguing asymmetry between
  the leading and trailing arm of the tidal tails.  The trailing arm grows
  faster than the leading arm.  This asymmetry is seen in the distance to
  first over-density and its size as well.  We establish a relation between
  the distance to the first over-density and the size of this over-density. 
\end{abstract}


\section{Introduction}
\label{sec:intro}

An image of the globular cluster Palomar~5 (Pal~5) and its tidal
tails has been published by \citet{Odenkirchen2001}.  In this image 
the tidal tails of Pal~5 along with some over-densities within them
are clearly visible.  Besides Pal 5, there are also other satellites
of the Milky Way (MW) which have elongated structures, for example the 
Sagittarius (Sgr) dwarf spheroidal galaxy (dSph), whose tidal tails
were described by  \citet{Majewski2003}, the Virgo stellar stream
\citep{Vivas2001} the new Aquarius stream \citep{Williams2011} and all
the streams listed in the Field of Streams \citep{Belokurov2006} . 

Tidal tails could be useful to constrain the properties of the Milky
Way (MW) like the mass and the shape of the dark matter (DM) halo.
\citet{Koposov2010}, using a 6D space map of the GD-1 stream in
Andromeda, found that an eccentric orbit in a flattened isothermal
potential (q$_{\Phi}$ = 0.87$^{+0.07}_{-0.04}$) fits the parameters of
distance, line-of-sight velocity and proper motion of the stars in the
stream.  \citet{Fellhauer2006} demonstrated that if the tails of the
Sagittarius dSph are wrapping around twice the MW, causing the visible
bifurcation \citep{Belokurov2006}, then the MW DM halo has to be
close to spherical.

To recover the Galactic potential using tidal tails, the first step is
to determine the orbit of the satellite by looking at the elongation
of its tails.  The next step is to perform simulations using different
Galactic potentials and assuming properties of the satellite like the
initial mass, the eccentricity of the orbit or the distance to its
apo-centre or peri-centre.  Finally, one has to compare the final
properties of the stars in the debris like the velocity dispersion,
line of sight velocity and proper motion with data from
observations.  The bias from this method comes from the fact that one
has to assume properties of the satellite and/or the Milky Way.
However, the study of the formation of tidal tails could give us some
clues about the initial properties of the disrupting satellites, thus
improving the methods to constrain Galactic properties. 

Following this line, \citet{Kupper2008} published a theory which
explains how tidal tails are formed.  In their work, they propose that
the stars that escape from a globular cluster move in epicycles inside
a galactic potential.  They found good agreement between their theory
and N-body simulations of globular clusters under the influence of a
point mass galaxy \citep{Kupper2008,Kupper2010,Kupper2012}.

In their theory, \citet{Kupper2008} make the approach that all
the particles escaping from the satellite have small velocities with
respect to the satellite and they escape always through the Lagrangian
points L1 and L2, therefore two symmetrical tails are formed, a leading
arm where particles orbit ahead of the satellite and a trailing arm where
particles orbit behind the satellite.  With these assumptions they
derive equations which predict the position and time of formation of
over-densities along both tails assuming that if D is the distance to
the first over-density in the leading arm, then -D is the distance to
the first over-density in the trailing arm where the reference frame is
in the centre of the satellite.

Our goal is to check the theory of \citet{Kupper2008} comparing
their predictions with simulations performed by SUPERBOX \citep[a
collision-less particle mesh code; see][]{Fellhauer2000} using a
dwarf satellite and modeling the Milky Way dark matter halo either
with a logarithmic halo (cored profile) or with a Navarro-Frenk-White
(NFW) halo \citep[cuspy profile][]{Navarro1996}.

The paper is organized as follows:  In Section~\ref{sec:setup} we
describe the properties of the satellites used in our simulations, the
dark matter potential and the orbits of the satellites.  In
Section~\ref{sec:analysis} we describe the analysis of the data and
how we estimate our errors.  In Section~\ref{sec:res} we present our
results.  In Section~\ref{sec:disc} we discuss the results and present
our conclusions.   

\section{Setup}
\label{sec:setup}

To integrate the particles forward in time, we use SUPERBOX
\citep{Fellhauer2000},  a collision-less particle mesh code which uses
3 levels of grids, with the two high-resolution grids
staying focused on the simulated and moving object.  This makes the
code faster by improving resolution only at the places of interest.  

To model the MW DM halo we use analytic potentials.  For the cored
case we use a logarithmic potential given by: 
\begin{eqnarray}
  \Phi & = & \frac{1}{2} v_{\rm c}^{2} \ln (r^2 + d^2),
\end{eqnarray}
where $r$ is the distance to the centre of the galaxy,
$v_{\rm c} = 220$~km\;s$^{-1}$ and $d = 12$~kpc.  

\begin{figure}
  \includegraphics[width=8.45cm]{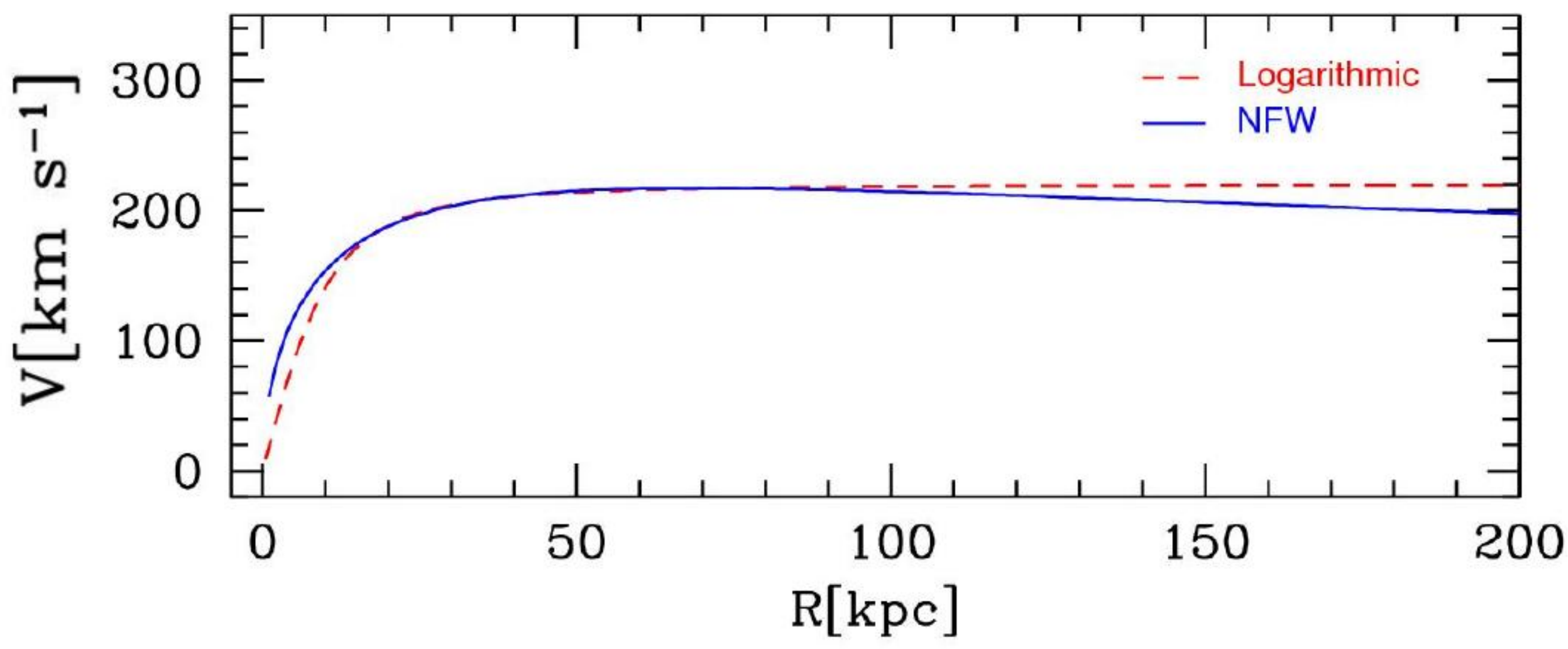}
  \caption{Rotation curves of the two potentials used to model the dark
    matter halo of the MW in our simulations.  The red dashed line is
    the rotation curve of the logarithmic profile and the solid blue
    line is the rotation curve of the NFW profile.  Note that both
    curves are very similar in the distance range we use to perform our
    simulations, in order to better compare the results.} 
  \label{fig:rot}
\end{figure}

Alternatively, to mimic a cusped potential, we use a
Navarro-Frenk-White potential: 
\begin{eqnarray}
  \Phi & = & 4 \pi G \rho_{0} \frac{r_{\rm s}^3}{r} \ln \left(
    \frac{r_{\rm s}}{r_{\rm s} + r} \right), 
\end{eqnarray}
with:
\begin{eqnarray}
 \rho_{0} = \frac{M_{200}}{4 \pi G r_{\rm s}^{3} (\ln (1+c) - c/(1+c))},
\end{eqnarray}
using $M_{200} = 1.8 \times 10^{12}$~M$_{\odot}$, $r_{200} = 200$~kpc
and $r_{\rm s}= 32$~kpc.  We have chosen these values to produce
similar rotation curves, which exhibit $v_{\rm rot,max}$ values as
expected for a galaxy like the MW (even though discs and bulges
are omitted).  The rotation curves for the two profiles can be seen
in Fig.~\ref{fig:rot}.  The inclusion of non-spherical
components and their influence on the tidal tails will be part of a
future study.

We investigate circular and elliptical orbits and the influence of the
satellite's mass in circular orbits, therefore we use 4 different
Plummer spheres \citep{Plummer1911}:
\begin{eqnarray}
  \label{eq:plum}
  \rho(r) & = & \frac{3{M}_{\rm pl}} {4\pi R_{\rm pl}^3} \left( 1 +
    \frac{r^2}{R^2_{\rm pl}}\right)^{-\frac{5}{2}}
\end{eqnarray}
with $M_{\rm pl}$ being the total mass and $R_{\rm pl}$ being the
scale-length of the Plummer profile, to model the satellites whose parameters
are given in Tab.~\ref{tab:plum}.  The relation between the values of $M_{\rm pl}$
and $R_{\rm pl}$ are chosen to have approximately the same filling factor
inside the tidal radius at a certain radius for the different
satellites.  The Plummer spheres are cut-off at $5 R_{\rm pl}$.

Now one can argue that, if we are using satellite galaxies, we should model
their dark matter halo as well.  The result would be that we would be
discussing invisible dark matter tails.  Instead we are using an one
component model in which mass follows light.  This could be identified by
a satellite galaxy which has lost its outer dark matter halo already or
a tidal dwarf galaxy, which was born without dark matter in the first place.

\begin{table}
  \centering
  \caption{ Description of the four different satellites used in the 
    simulations.  Column 1 is the number to identify each satellite,
    column 2 and 3 are the number of particles and the mass of the
    Plummer sphere and column 4 and 5 are the scale radius of the
    Plummer profile and the half mass radius of the satellite.} 
  \label{tab:plum}
  \begin{tabular}{|c|c|r|l|l|}
    \hline
    & $N$ & $M_{\rm pl}$ &  $R_{\rm pl}$ & r$_{\rm h}$ \\
    &     & [M$_{\odot}$]& [kpc]       & [kpc]     \\   
    \hline 
    1 & $10^6$ & $0.5 \times 10^8$ & $0.2$ & $0.34$ \\  
    2 & $10^6$ & $10^8$          & $0.35$ & $0.595$ \\
    3 & $10^6$ & $2 \times 10^8$ & $0.45$ & $0.765$ \\
    4 & $10^6$ & $4 \times 10^8$ & $0.6$  & $1.02$ \\
    \hline
\end{tabular}
\end{table}

\begin{table}
  \centering
  \caption{ Velocity of satellite 2 at five different distances from
    the galactic centre to get circular orbits in the logarithmic
    potential (second column) and NFW potential (third column).} 
  \label{tab:dist}
  \begin{tabular}{|c|l|l|}
    \hline
    $R$ & $v_{\rm log}$ & $v_{\rm NFW}$ \\
     $[$kpc$]$ & [km\;s$^{-1}$] & [km\;s$^{-1}$] \\
    \hline
    $15$ & $171.79$ & $174.36$ \\
    $20$ & $188.65$ & $187.75$ \\
    $25$ & $198.34$ & $196.90$ \\
    $30$ & $204.27$ & $203.34$ \\
    $35$ & $208.11$ & $207.93$ \\
    $50$ & $218.43$ & $214.66$ \\
    \hline
  \end{tabular}
\end{table}

\begin{table}
  \centering
  \caption{Eccentricity of the orbit, velocity of the satellite at
    apo-galacticon ($80$~kpc) and peri-galactic distance for the
    eccentric orbits used in this study.}
  \label{tab:ecc}
  \begin{tabular}{|c|c|c|}
    \hline
    $e$ & $v_{\rm apo}$ & $r_{\rm peri}$ \\
        & [km\;s$^{-1}$] & [kpc] \\
    \hline
    $0.1$ & $195.64$ & $65.45$ \\
    $0.2$ & $174.21$ & $53.33$ \\
    $0.3$ & $153.07$ & $43.08$ \\
    $0.4$ & $132.05$ & $34.29$ \\
    $0.5$ & $110.96$ & $26.67$ \\
    $0.8$ & $ 45.21$& $ 8.89$ \\
    \hline
  \end{tabular}
\end{table}

We place satellite~2 (as standard model) at $6$ different distances from
the centre of the galaxy according to Tab.~\ref{tab:dist} and give them
velocities to produce circular orbits and let them evolve for $5$~Gyr
in both dark matter halos.

Then we investigate the influence of eccentricity, so we keep a fixed
apo-galacticon of $80$~kpc and change the velocity of the satellite 2
according to Tab.~\ref{tab:ecc} to produce elliptical orbits with
eccentricities between $0.1$ and $0.8$. 

Finally, we investigate the influence of the satellite's mass
using the logarithmic halo only and satellites with a mass of $0.5$,
$1.0$, $2.0$ and $4.0 \times 10^8$~M$_{\odot}$ orbiting at $25$~kpc for
$5$~Gyr. 

\section{Analysis}
\label{sec:analysis}

To analyse the tidal tails we use a code developed by \citet{Vejar2013}.
This code divides both tails into bins of equal size and counts the
number of stars in each bin.  The size of each bin is given by the
maximum length of tails that we want to analyse ($L$) and the number of
bins we want to use ($N_{\rm bin}$), therefore each bin will have a
size equal to $L_{\rm bin} = L / N_{\rm bin}$.  To count the number of
stars in each bin, the code finds the centre of density of the
satellite at each time-step from an output file of SUPERBOX containing
this information.  Then the code calculates the angular position of
the satellite's centre of density and the angular position of each star
with respect to the $x$-axis, where the reference frame is centered 
on the center of the galactic potential and the initial position of the
satellite is on the $x$-axis.  Then an array containing this information
for each particle is sorted by its angular position.  The code puts
the centre of the first bin in the centre of density of the satellite
and calculates the distance between the particles and the centre of
the bin using the law of cosines.  If the distance is smaller than half
the size of the bin, then the particle is counted in that bin.

\begin{figure}
  \includegraphics[width=8.4cm,height=8.4cm]{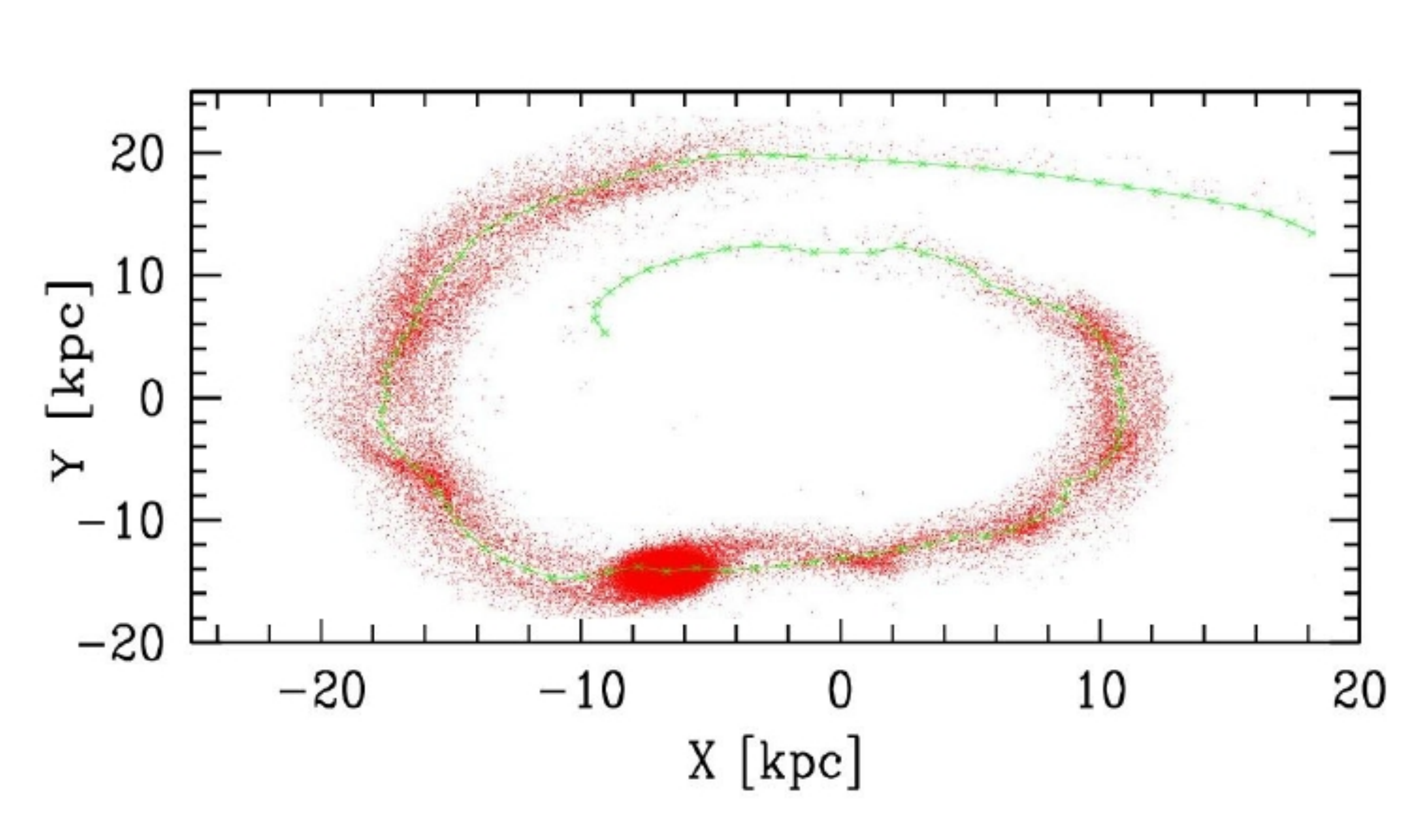}
  \caption{Path followed by the code (green line) to analyse tidal tails
    produced in a satellite orbiting at $15$~kpc from the galactic centre} 
  \label{fig:rvejarbins}
\end{figure}

To find a suitable place where to put the centre of the next bin the
code calculates the average positions $\hat{x}$, $\hat{y}$ and
$\hat{z}$ of the particles outside the bin.  Then the distance
$\hat{r}$ between the centre of the previous bin and the average
position of the particles outside the bin is calculated using the
Pythagoras Theorem.  If $\hat{r} / L_{\rm bin} = 1$ then we have a
place to set the centre of the next bin.  This procedure is done until 
the code reaches the maximum length $L$ defined by the user as we
can see in Fig.~\ref{fig:rvejarbins}.

Once we have the data of the number of particles in each bin along
both tails we find that some bins have a number of particles larger
than the average, this are the bins where the over-densities are
located.  However, identifying over-densities only by the number of
particles is not very accurate, so we use another criterion.  To
detect over-densities properly we pick pieces of $10$~kpc of tails and
calculate the average number of particles in each bin inside those
$10$~kpc, then we detect an over-density by selecting the bins where the
number of particles is larger than the average.  By doing this we have
the information of the distance to each bin which carries an
over-density of stars.  Usually there are several bins in an
over-density, therefore, to measure the distance to an over-density we
take the distance to the bin which is in the centre of the
over-density.  We also define the size of an over-density as the
distance to the last bin in the over-density minus the distance to the
first bin in the over-density. 

As the measurements of distance and size depends on the size of the
bins, used to analyse the tails, we measure this quantities using
bins of $0.4$~kpc, $0.5$~kpc and $0.6$~kpc and take the mean value
in each case, then we can also calculate the error.  The results of
our simulations show that by choosing smaller bin-sizes, the results
become unreliable because of low number statistics and larger
bin-sizes will smear out the features.

After counting stars in each bin at each time-step, the code presents
the information in a colour density plot, in the $x$-axis is the time in
intervals of the time-step chosen for the output of the simulation
data ($10$~Myr in our case) in the $y$-axis it shows the longitudinal
separation: at the centre ($0$~kpc) is the satellite, at positive
distances is the leading arm and negative distances represent the
trailing arm.  The colour scheme represents the logarithmic number
densities of the tails. 

\begin{figure}
  \includegraphics[width=240px]{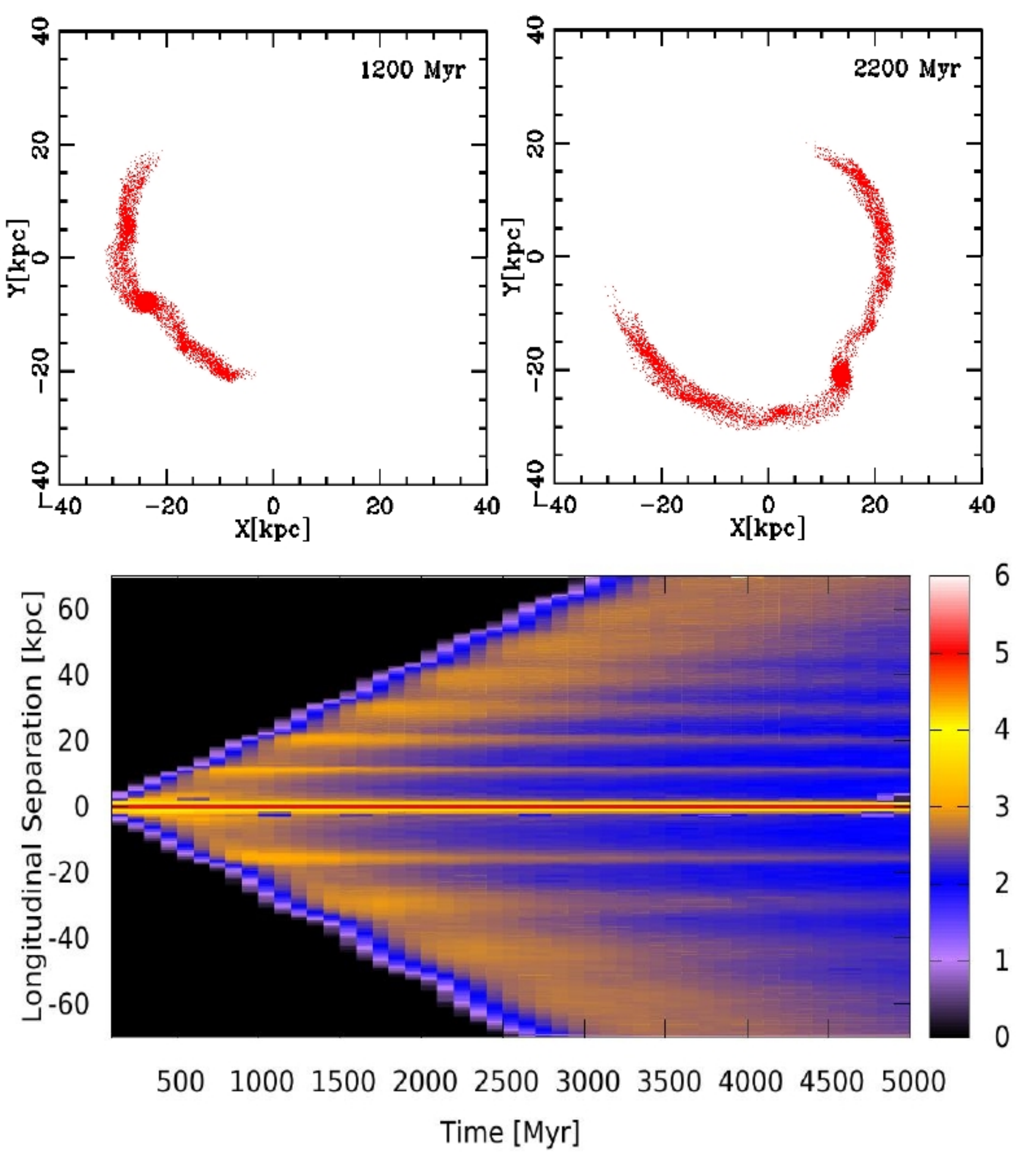}
  \caption{Lower panel: Colour density plot of the number of stars
    along the tails for a dSph orbiting a logarithmic dark matter halo
    for $5$~Gyr on a circular orbit in intervals of $100$~Myr.  The colour
    bar shows the logarithm of the particles per bin.
    Particles in the leading arm are assigned positive distances
    whereas particles in trailing arm have negative distances.  In the
    top left panel we have the satellite and its tidal tails at
    $1.2$~Gyr and in the right panel the same satellite at $2.2$~Gyr.} 
  \label{fig:example}
\end{figure}

We can see two different evolutionary stages of satellite~2 orbiting
at $25$~kpc from the Galactic Centre in the cored halo in
Figure~\ref{fig:example}.  After using the analysing code described in
Sect.~\ref{sec:analysis} we obtain the colour plot shown in the bottom
panel.  We clearly see how the tails grow with time.  Furthermore, we
see as yellow horizontal lines the over-densities which develop with
time and which stay at constant distances to the satellite with time
(on circular orbits).

\section{Results}
\label{sec:res}

Once we have the data of the position and time of formation of
over-densities we can compare our results with the expected values from
the theory. 

According to \citet{Kupper2008} the equations of motion for a
particle that escapes from the tidal radius of a satellite with small
velocity with respect to the satellite are given by: 
\begin{eqnarray}
  \label{eq:kuepper}
  x & = & \frac{4 \Omega ^2}{\kappa ^2}x_{\rm L} + \left( 1 - \frac{4
      \Omega ^2}{\kappa ^2}   \right) x_{\rm L} \cos \kappa t, \\
   \label{eq:kuepper-y}   
  y & = & - \frac{2 \Omega}{\kappa} \left( 1 - \frac{4 \Omega
      ^2}{\kappa ^2} \right) x_{\rm L} ( \sin \kappa t - \kappa t ), \\
  z & = & 0.
\end{eqnarray}
With the reference frame in the centre of the satellite, the $x$-axis
points towards the Galactic anti-centre and the $y$-axis points in the
direction of the motion of the satellite.  $\kappa$ is the epicyclic
frequency, $\Omega$ is the angular velocity of the satellite and
$x_{\rm L}$ is the tidal radius of the satellite which we calculate
according to Eq.~\ref{eq:tidal}, $r$ is the radius of the
orbit, $M$ is the mass of the satellite and $M(r)$ is the 
mass of the halo enclosed inside a radius $r$.  $\kappa$ is calculated
according to Eq.~\ref{eq:kappa}: 
\begin{eqnarray}
  \label{eq:tidal}
  x_{\rm L} & = & r \left( \frac{M}{3 M(r)}\right)^{1/3}, \\
  \label{eq:kappa}
  \kappa ^2 & = & \frac{\partial ^2 \Phi}{\partial r^2} + 3\Omega ^2.
\end{eqnarray}
If we assume a point-mass potential for the galaxy, then $\kappa =
\Omega$.  These equations are based on the standard potential- and
epicyclic- theories \citep[see e.g.][]{GalDyn}.

Escaped particles start to de-accelerate at the turn-around points of
their epicyclic orbits, i.e.\ when they perform a complete oscillation
of 2$\pi$ at $t = t_{\rm cp} = 2 \pi / \kappa$ therefore if many
particles escape from the satellite and decelerate at the same
position, we will see an over-density of stars at a distance $y(t_{\rm
  c}) = y_{\rm cp} = 12 \pi x_{\rm L}$ from the centre of the satellite. 

The predicted values of $t_{\rm c,p}$ and $y_{\rm c,p}$ from the point
mass-approximation as well as if we use the real potential from our
simulations $t_{\rm c}$ and $y_{\rm c}$ are presented in Tabs.~\ref{tab:log}
\& \ref{tab:nfw} along with the values obtained from our simulations
($t_{\rm s}$ and $y_{\rm s}$). 

If we do not assume a point-mass galaxy, then we need
to solve Eq.~\ref{eq:kuepper-y} for $t = t_{\rm c} = 2 \pi / \kappa$.
Therefore the position of the first over-density along the 
y-axis and the time at which is formed are given by:
\begin{eqnarray}
  \label{eq:yc}
  y_{\rm c} & = &  \frac{4 \pi \Omega}{\kappa} \left( 1 - \frac{4 \Omega
      ^2}{\kappa ^2} \right) x_{\rm L}, \\
  t_{\rm c} & = & \frac{2 \pi}{\kappa}.    
\end{eqnarray}

\begin{table*}
  \centering
  \caption{Predicted values of distance and time of formation of the
    first over-density for leading arm according to the theory of
    \citet{Kupper2008} for a point-mass galactic
    potential($t_{\rm c,p}$ \& $y_{\rm c,p}$) (columns 2 \& 3), assuming
    a Logarithmic MW potential ($t_{\rm c}$ \& $y_{\rm c}$) (columns 4
    \& 5), the time of formation and position of the first over-density
    from our logarithmic halo simulations $t_{\rm s}$ and $y_{\rm s}$ (columns 6 \& 7).} 
  \label{tab:log}
  \begin{tabular}{|r|r|r|r|r|r|r|}
    \hline
    $D$ & $t_{\rm c,p}$ & $y_{\rm c,p}$ & $t_{\rm c}$ & $y_{\rm c}$ &
    $t_{\rm s}$ & $y_{\rm s}$ \\ 
    (kpc) & [Myr] & [kpc] & [Myr] & [kpc] & [Myr] & [kpc] \\
    \hline
    $15$ & $536.798$ & $38.961$ & $322$ & $3.416$ & $477 \pm 24$ &
    $5.863 \pm 0.285$ \\
    $20$ & $651.774$ & $44.342$ & $410$ & $5.403$ & $580 \pm 10$ &
    $8.620 \pm 0.430$ \\
    $25$ & $774.927$ & $49.765$ & $503$ & $7.369$ & $647 \pm 38$ &
    $11.301 \pm 0.133$ \\
    $30$ & $902.917$ & $55.104$ & $599$ & $9.224$ & $760 \pm 26$ &
    $13.700 \pm 0.254$ \\ 
    $35$ & $1033.950$ & $60.314$ & $695$ & $10.949$ & $883 \pm 18$ &
    $15.839 \pm 0.299$ \\ 
    $50$ & $1436.906$ & $75.111$ & $989$ & $15.460$ & $1243 \pm 27$ &
    $20.610 \pm 0.359$ \\ 
    \hline
  \end{tabular}
\end{table*}

\begin{table*}
  \centering
  \caption{Predicted values of distance and time of formation of the
    first over-density for leading arm according to the theory of
    \citet{Kupper2008} for a point-mass galactic
    potential($t_{\rm c,p}$ \& $y_{\rm c,p}$) (columns 2 \& 3), assuming
    a NFW MW potential ($t_{\rm c}$ \& $y_{\rm c}$) (columns 4
    \& 5), the time of formation and position of the first over-density
    from our NFW simulations $t_{\rm s}$ and $y_{\rm s}$ (columns 6 \& 7).} 
  \label{tab:nfw}
  \begin{tabular}{|r|r|r|r|r|r|r|}
    \hline
    $D$ & $t_{\rm c,p}$ & $y_{\rm c,p}$ & $t_{\rm c}$ & $y_{\rm c}$ &
    $t_{\rm s}$ & $y_{\rm s}$ \\ 
    (kpc) & [Myr] & [kpc] & [Myr] & [kpc] & [Myr] & [kpc] \\
    \hline
    $15$ & $528.896$ & $38.577$ & $331$ & $4.517$ & $450 \pm 10$ &
    $7.196 \pm 0.323$ \\
    $20$ & $654.910$ & $44.484$ & $417$ & $5.872$ & $543 \pm 15$ &
    $9.974 \pm 0.110$ \\
    $25$ & $780.564$ & $50.006$ & $505$ & $7.293$ & $623 \pm 38$ &
    $11.484 \pm 0.084$ \\
    $30$ & $907.017$ & $55.271$ & $596$ & $8.772$ & $763 \pm 18$ &
    $13.524 \pm 0.093$ \\ 
    $35$ & $1034.834$ & $60.349$ & $688$ & $10.302$ & $900 \pm 15$ &
    $15.619 \pm 0.297$ \\ 
    $50$ & $1428.756$ & $74.827$ & $981$ & $15.144$ & $1277 \pm 38$ &
    $20.270 \pm 0.407$ \\ 
    \hline
  \end{tabular}
\end{table*}

We are analysing the simulations using satellite 2 on circular orbits at different
distances from the galaxy in both analytic potentials with respect to the
location and formation time of the first over-density and compare our results
with the prediction formulas derived above.  The values are displayed in
Tabs.~\ref{tab:log} and~\ref{tab:nfw}.

Using the point-mass approximation the time of formation of the first
over-density is over-predicted by a factor of $\approx 1.2 \pm 0.1$.  If we use
the equations for the real potential the situation reverses and we
under-predict the time of formation by a factor of $\approx 0.7 \pm 0.1$.  This
is true irrespective of the potential (logarithmic or NFW) we are using.

A similar picture presents the predictions for the location (i.e. distance) of the
first over-density.  Again the point-mass approximation over-predicts the location,
now by a factor ranging between $6.6$ and $3.6$ with a decline roughly $\propto 1/r$.
The full equations again deliver an under-prediction of the simulation values ranging
from $0.58$ and $0.75$.  Again we see a $1/r$ dependence of the results, i.e.\ the
further away we are from the centre of the galaxy the closer simulation values and
their respective predictions get.

We conclude that the point-mass approximation is not suitable to predict simulation
results using more realistic galactic potentials.  That the predictions for the
formation time, using the real potential, are somewhat smaller than the measured
values can be explained that the theory gives the time when the first stars reach the
point of the over-density and we need some time more to establish a detectable
signal (i.e.\ over-density).  This was already pointed out in \citet{Kupper2008}.

For the mismatch of the position one could imagine that it is caused by the more
extended object we are using in this study and that the Lagrangian points of escape
are no longer almost coinciding with the distance of the orbit, i.e.\ that $\Delta r
= r \pm x_{\rm L}$ is not negligible any longer.  But this would mean that
we see over- and under-predictions depending if we analyse the leading or the trailing
arm of our object.  This is not the case in our simulations.  We only detect a
smaller deviation for the leading arm than for the trailing arm but both directed in
the same direction with respect to over- and under-predictions.  We suspect
that the zero velocity assumption (i.e.\ that stars leave the satellite with small
velocities only) is no longer valid for the extended objects we use. 

The discrepancy between the point-mass approximation and the real simulation values
was not visible in the results of \citet{Kupper2008}, as they were actually using a
point-mass potential to mimic the galaxy together with the point-mass approximation
to compare their results.  In that sense we agree with \citet{Kupper2008}, just
one has to use the prediction equations matching the used potential.

After comparing the theory with our simulations, we analyse the
rate of growth of the tails arising from the satellites.  As the growth
of the tails is linear with time for satellites on circular orbits, we
simply take two values of the length of the tails together with 2 values of
time and calculate the slope.  We use the length at $3000$~Myr and the
length at $0$~Myr, which is zero to calculate the growth-rate.
For satellites orbiting on eccentric orbits, the growth of the tails
is not linear with time, tails are stretched and compressed as the
satellite passes through peri- and apo-galacticon, respectively.  To
measure the rate of growth we have taken the values for the length of
the tails between the time after two consecutive peri-centre passages
measured half-way to the next apo-galacticon.

\begin{figure}
  \includegraphics[width=8.4cm]{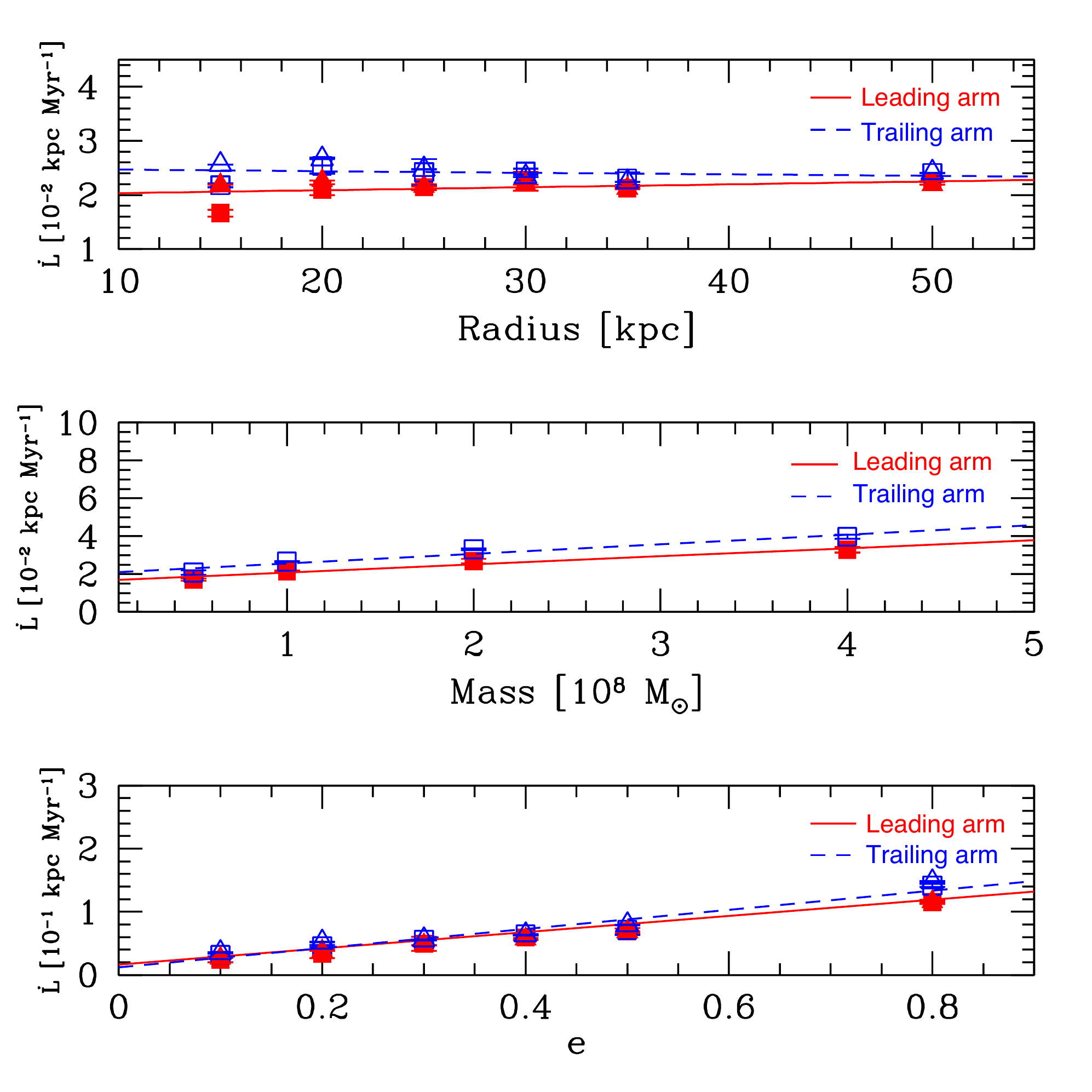}
  \caption{Rates of growth of leading and trailing arm depending on
    the radius of circular orbits (top) the mass of the satellite
    (middle) and the eccentricity of the orbit (bottom).  We plot data
    from simulations using a logarithmic halo as squares and a NFW profile
    as triangles.  Red filled symbols are data from the leading arm and
    blue open symbols are data from the trailing arm.  Red and blue lines
    are the best linear fits for leading and trailing arm respectively.
    As we see no difference between the logarithmic and NFW halo results
    we fit both results simultaneously.  The exception is the middle panel
    as for varying satellite mass we only used the logarithmic potential.} 
  \label{fig:growth}
\end{figure}

We find that the trailing arm always grows faster than the leading arm
irrespective of circular or elliptical orbits (see Fig.~\ref{fig:growth}).
For circular orbits, the rates of growth are independent of the radius of
the orbit and $\sim 0.02$~kpc\;Myr$^{-1}$ for leading arms and
$\sim 0.025$~kpc\;Myr$^{-1}$ for trailing arms as shown in the top panel
of Fig.~\ref{fig:growth}.  One can also see that these results are
independent of the form of the potential, i.e.\ using a logarithmic or
a NFW halo.

This result shows that for the growth of the tidal tails, the underlying
potential is of no importance.  The velocity of the tail growth is
given by the parameters of the satellite only.  The fact that the trailing
arm is growing slightly faster can be understood by the fact that stars
are leaving with similar velocities through the Lagrangian points.  For
an extended object as used in our simulations, those stars have slightly
different angular velocities compared to that of a circular orbit at this
distance.  These differences could different in magnitude for the leading
and trailing arm leading to this asymmetry of tail growth.

In the middle panel of Fig.~\ref{fig:growth} we see clearly that the growth
depends on the mass of the satellite in a linear fashion.  The rate of growth
for the the leading arm varies from 
$0.0172 \pm 0.0004$ ~kpc\;Myr$^{-1}$ for a satellite of $0.5 \times
10^8$~M$_{\odot}$ up to $0.0329 \pm 0.0009$~kpc\;Myr$^{-1}$ for a
satellite of $4.0 \times 10^8$~M$_{\odot}$.  The best linear fits are
given by the Eq.~\ref{eq:growratemasslead} for the leading arm and
Eq.~\ref{eq:growratemasstrail} for the trailing arm:
\begin{eqnarray}
  \label{eq:growratemasslead}
  \dot{L_{\rm L}} & = &  0.0043 \pm 0.0007 \times M + 0.017 \pm 0.002 , \\
  \label{eq:growratemasstrail}
  \dot{L_{\rm T}} & = &  0.0051 \pm 0.0010 \times M + 0.021 \pm 0.002 .
\end{eqnarray}
For these simulations we have only used the logarithmic potential as
galactic potential.

Here we clearly see the dependence on the internal satellite parameters.
Stars leaving the satellite should have escape velocity or slightly higher.  
We conclude that the assumption that stars leave the satellite with zero or
negligible relative velocity may be valid for small star clusters but not
for very massive and extended objects.  In this respect our study shows
a disagreement with the assessment of \citet{Kupper2008}.

Finally, we find that the rate of growth of the tails measured between
peri-galacticon and the next apo-galacticon for satellites orbiting in
eccentric orbits seems to depend linearly on the eccentricity of the
orbit and varies from  $0.0241 \pm 0.0022$~kpc\;Myr$^{-1}$ for the
leading and $0.0320 \pm 0.0013$ for trailing arm for an eccentricity of
$0.1$, up to $0.1157 \pm 0.0019 $ and $0.1474 \pm 0.0063$~kpc\;Myr$^{-1}$
for leading and trailing arm, respectively, for an eccentricity of $0.8$.

The best linear fits are shown in Eq.~\ref{eq:growrateclead} for the rate
of growth of the leading arm and in Eq.~\ref{eq:growratectrail} for the
rate of growth of the trailing arm.
\begin{eqnarray}
  \label{eq:growrateclead}
  \dot{L_{\rm L}} & = &  0.128 \pm 0.005 \times e + 0.010 \pm 0.002 , \\
  \label{eq:growratectrail}
  \dot{L_{\rm T}} & = &  0.152 \pm 0.013 \times e + 0.012 \pm 0.006 .
\end{eqnarray}
In our simulations this is explained easily.  The higher the eccentricity
the deeper into the galactic potential the satellite is orbiting.  Stars
get lost by tidal shocks at peri-galacticon passages.  The closer to the
galactic centre the satellite is orbiting the smaller the instantaneous
tidal radius will be and the more stars can get stripped, which have higher
relative velocities with respect to the satellite.  Again here we see
a clear deviation from the zero velocity assumption.

\begin{figure}
  \includegraphics[width=8.4cm]{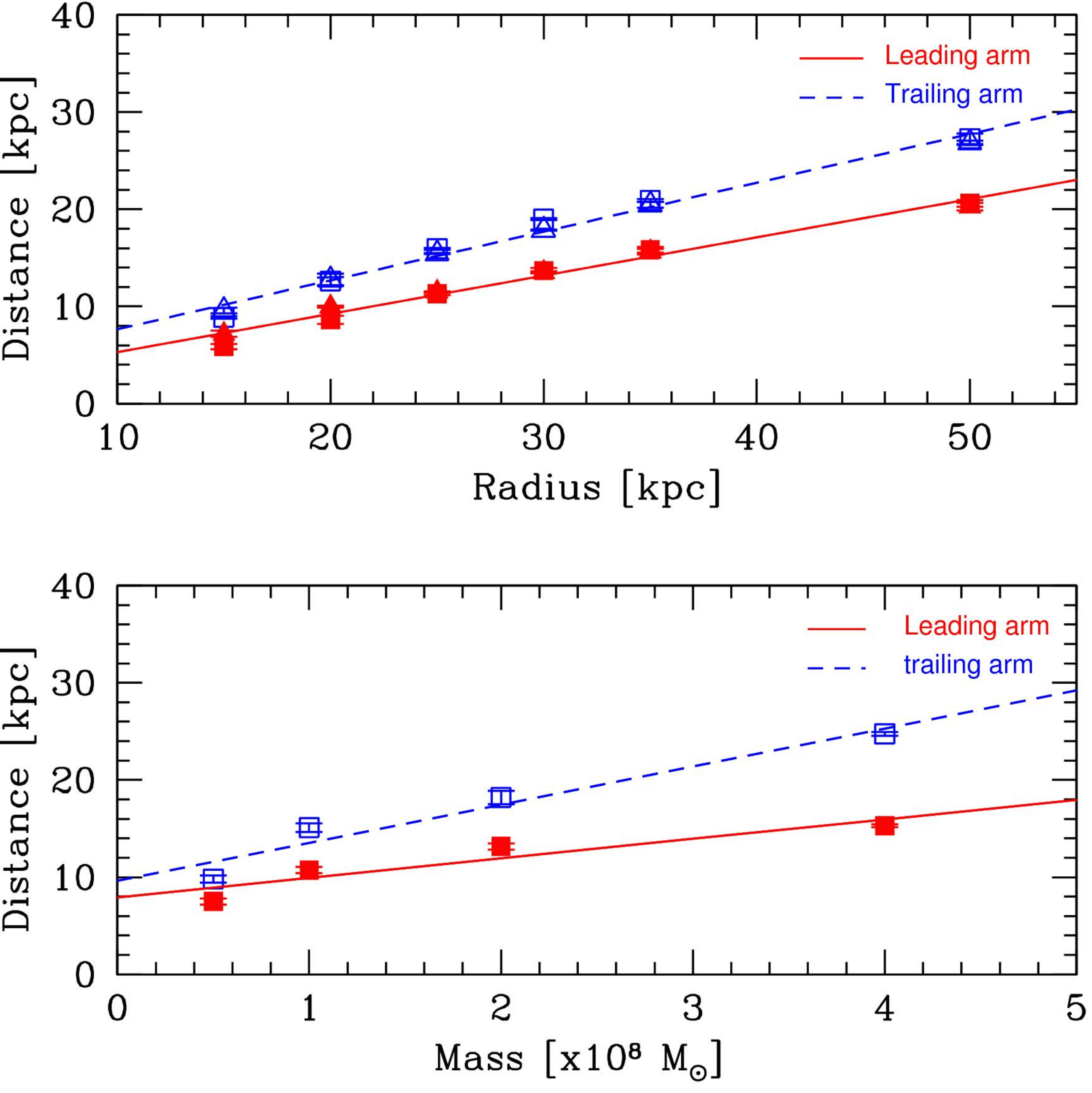}
  \caption{Top panel: Distance to the first over-density in leading and
    trailing arms as function of the radius of the orbit.  Bottom
    panel: Distance to the first over-density as function of the
    satellite mass.  We plot data from simulations using a logarithmic halo
    as squares and a NFW profile as triangles.  Red filled symbols are data
    from the leading arm and blue open symbols are data from the trailing arm.
    Red and blue lines are the best linear fits for leading and trailing arm
    respectively.  As we see no difference between the logarithmic and NFW halo
    results we fit both results simultaneously.  Again in the lower panel we
    only show simulations using the logarithmic halo for the varying satellite
    mass.}
  \label{fig:dist}
\end{figure}

In Fig.~\ref{fig:dist} we show the distance to the first over-density in
the leading and the trailing arm.  We find that the distance to the first
over-density in leading and trailing arm depends on the radius of the
orbit and the mass of the satellite as expected from the theory of
\citet{Kupper2008} but does not depend on the potential used to
model the MW.  This can be seen in the top panel as the symbols for the
logarithmic halo and the NFW profile basically overlap.

We also find that in the leading arm the first over-density is always at a
smaller distance from the satellite than in the trailing arm.  After the
formation of the over-densities, the distance to them remains constant
(see example in Fig.~\ref{fig:example}).

The difference in the distance between the two tails is a puzzling detail
of our simulations.  It could be explained by the fact that we have a much
more extended object than used in the previous study.  Again, the
difference in distance between the two Lagrangian points should be taken
into account.  But, then we would expect that this difference is more
pronounced the closer we are to the centre of the galaxy.  That we see
the opposite trend in our results is therefore counter-intuitive and
needs further investigation.  Nevertheless, it is a fact which should be
taken into account when analysing observational results.  The fact
that tidal tails of one and the same object have not same length and that the
over-densities appear at different locations is not
necessarily a sign for an interaction with another object.

We have studied the distance to the first over-density in the leading
and trailing arm.  In reality, we have measured the distance to the
centre of the over-density, because the over-densities have a size. 
We calculate the mean value of all bins which present the
over-density.  As the size of these over-densities may help us to
distinguish further between Galactic and/or satellite properties, we
now take a closer look at the sizes. 

\begin{figure}
  \includegraphics[width=8.4cm]{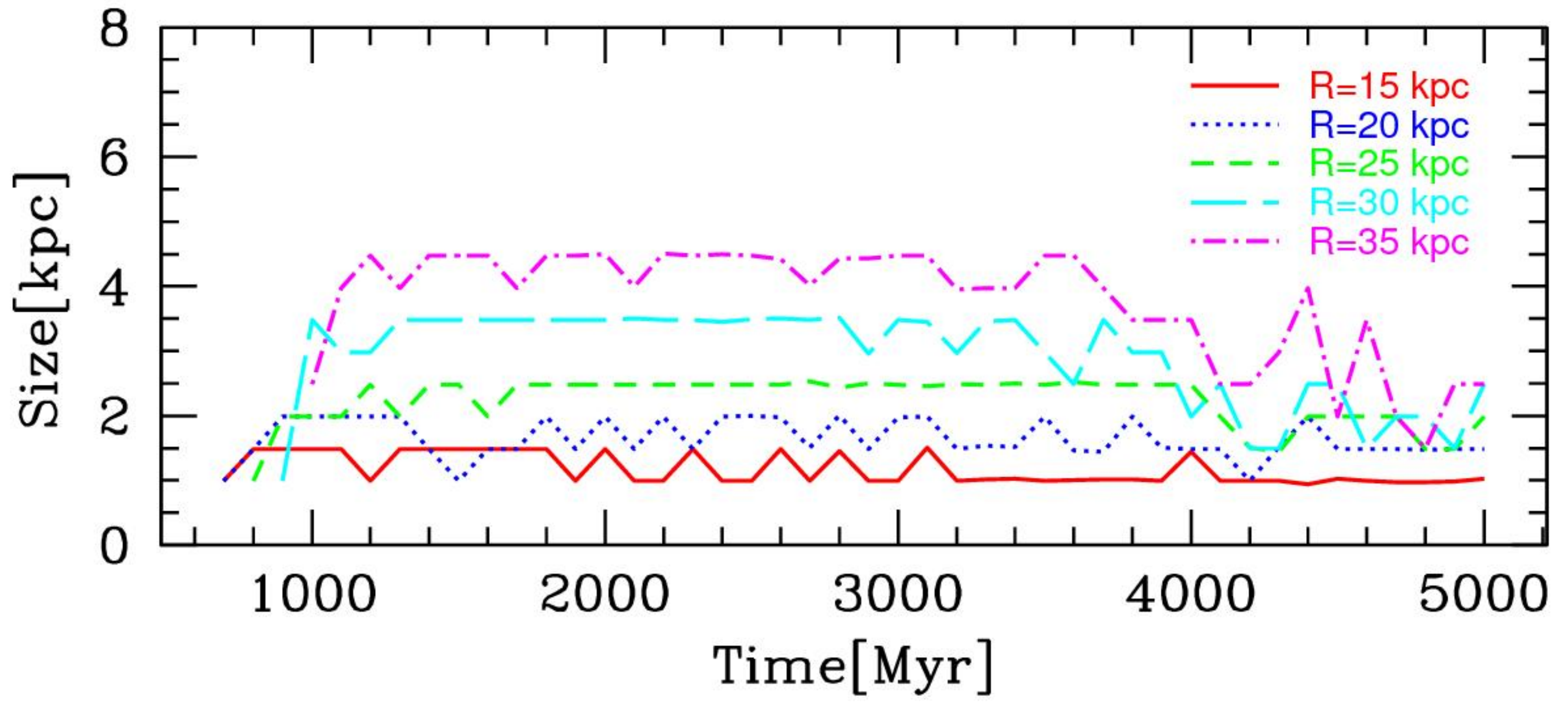}
  \caption{Size of the first over-density as function of time for
    a satellite orbiting at 15 kpc (solid red line), 20 kpc (blue dotted line), 
    25 kpc (green short dashed line), 30 kpc (cyan long dashed line) and 35 kpc
    (magenta dot dashed line). We can note that after $\sim$ 3600 Myr the
    size of the over-density is decreasing down to about 2 kpc.} 
  \label{fig:sizetrail}
\end{figure}

\begin{figure}
  \includegraphics[width=8.4cm]{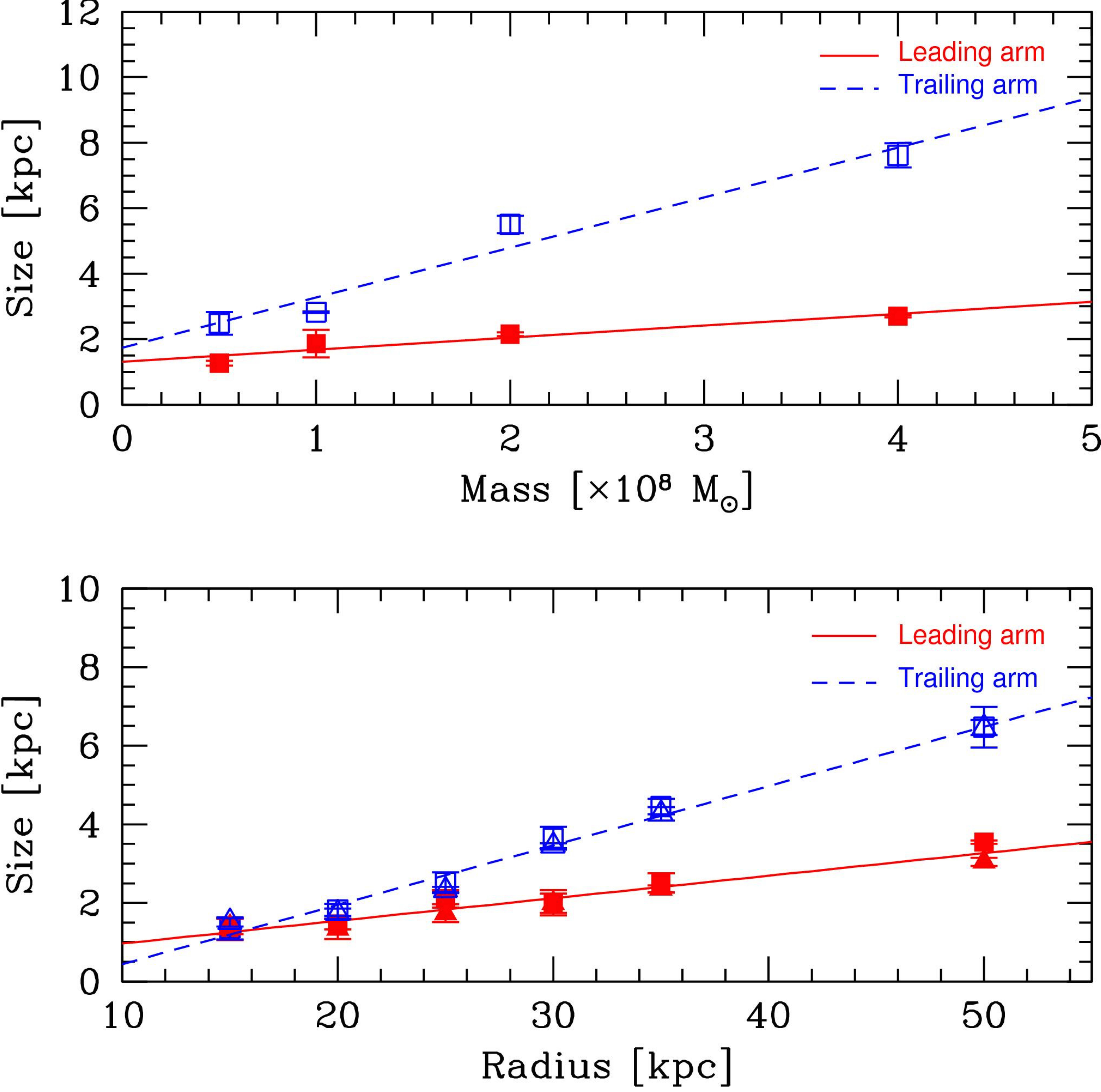}
  \caption{Size of the first over-density as function of the mass of
    the satellite (top panel).  In the lower panel we plot the
    size of the first over-density as function of the radius of the
    orbit.  We plot data from simulations using a logarithmic halo
    as squares and a NFW profile as triangles.  Red filled symbols are data
    from the leading arm and blue open symbols are data from the trailing arm.
    Red and blue lines are the best linear fits for leading and trailing arm
    respectively.  As we see no difference between the logarithmic and NFW halo
    results we fit both results simultaneously.  Again in the top panel we
    only show simulations using the logarithmic halo for the varying satellite
    mass.}
  \label{fig:size}
\end{figure}

We find that after the formation of the over-densities they have a
nearly constant size, however, the size is decreasing after $\sim
3600$~Myr as we can see in Fig.~\ref{fig:sizetrail}  

To study how the size depends on the properties of the satellite we
take an average of the size of the over-densities between $1$ and
$3$~Gyr, i.e., a time-span, starting after the formation of the
over-densities, where the size of them does not vary significantly.
The results are shown in Fig.~\ref{fig:size}.

We find that the size of the first over-density in the leading arm
depends only weakly on the radius of the orbit and does not depend on
the potential used to model the MW.  The size varies from $1.33 \pm 0.16$~kpc
for a satellite orbiting at $15$~kpc up to $3.55 \pm 0.05$~kpc for a satellite
orbiting at $50$~kpc.  In the trailing arm we see a clear dependence
of the size of the first over-density with the radius of the orbit.
The size of the first over-density increases with increasing radius of
the orbit from $1.35 \pm 0.16$~kpc for a circular orbit at $15$~kpc up to
$6.47 \pm 0.30$~kpc for a circular orbit at $50$~kpc.

The size of the first over-density depends on the mass of the
satellite.  In the trailing arm the size of the first over-density
increases from $2.49 \pm 0.20$~kpc for a satellite with a mass of $0.5 \times
10^8$~M$_{\odot}$ up to $7.62 \pm 0.22$~kpc for a satellite with a mass of $4.0
\times 10^8$~M$_{\odot}$.  In the leading arm, the size of the first
over-density depends only weakly on the mass of the satellite ranging
from $1.27 \pm 0.04$~kpc for a satellite with a mass of $0.5 \times
10^8$~M$_{\odot}$ up to $2.71 \pm 0.02$~kpc for a satellite with a mass of $4.0
\times 10^8$~M$_{\odot}$.  Apparently, the size of the first
over-density in leading arms depends linearly with the mass of the
satellite, however, for trailing arms it could be a power law. 
We show in Fig.~\ref{fig:size} the best linear fits to the data
of the size of the first over-density in the leading arm and trailing arm
depending on the radius of the orbit in kpc (Eq.~\ref{eq:sizeradiuslead} 
and Eq.~\ref{eq:sizeradiustrail}) and the Mass of the satellite in M$_{\odot}$ 
(Eq.~\ref{eq:sizemasslead} and Eq.~\ref{eq:sizemasstrail}):
\begin{eqnarray}
  \label{eq:sizeradiuslead}
  S_{\rm L} & = &  0.058 \pm 0.005 \times r + 0.4 \pm 0.1 , \\
  \label{eq:sizeradiustrail}
  S_{\rm T} & = &  0.15 \pm 0.01 \times r - 1.1 \pm 0.2 . \\
  \label{eq:sizemasslead}
  S_{\rm L} & = &  0.37 \pm 0.08 \times M + 1.3 \pm 0.2 , \\
  \label{eq:sizemasstrail}
  S_{\rm T} & = &  1.5 \pm 0.2 \times M + 1.7 \pm 0.5 .
\end{eqnarray}

Again we see a clear asymmetry of the two tails.  The size of the first over-density
in the trailing arm grows about three times faster with mass and distance than the
one of the leading arm.  This cannot be explained by the use of an extended object
alone.  Even though we do not have a valid explanation for this finding, it is a
fact which should be taken into account when observing real objects.

\begin{figure}
  \includegraphics[width=8.4cm]{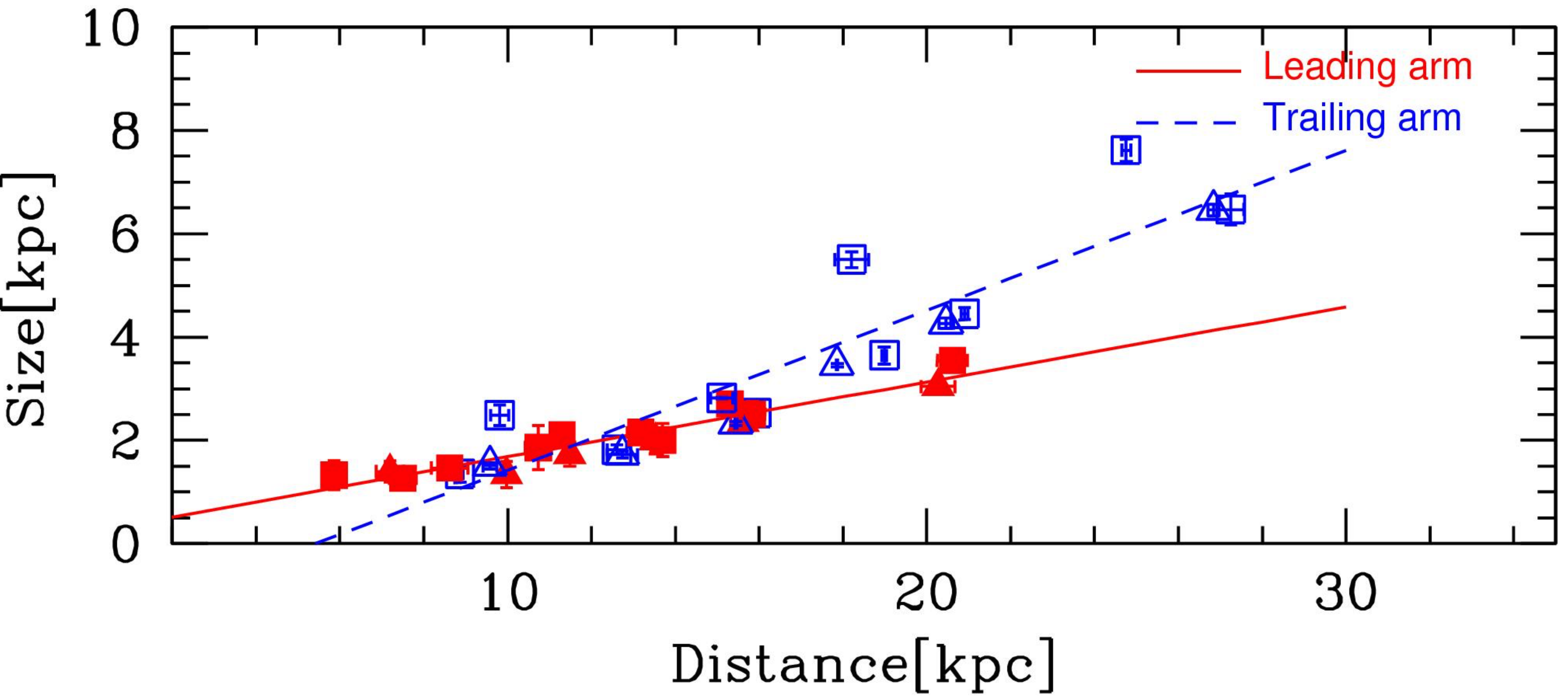}
  \caption{Size of the first over-density as function of the distance to
    the first over-density for trailing arms (blue symbols) and leading
    arms (red symbols) using simulations of a satellite orbiting a
    logarithmic halo on circular orbits and the best linear fits in red
    for the leading and blue for the trailing arm.} 
  \label{fig:combine}
\end{figure}

Finally, we combine the data from the distance to the first over-density
in the leading and trailing arms with the size of the first
over-density for simulations of satellites orbiting a logarithmic halo
in circular orbits.  The data is shown in Fig.~\ref{fig:combine}.
The reason for this exercise is that in this way we might be able to
distinguish between internal and external effects, i.e.\ mass of the
satellite and the potential strength.  Unluckily, we do not see such a
trend in Fig.~\ref{fig:combine} as all data-points follow the same relation,
no matter the mass of the satellite or the distance to the galaxy.  

Nevertheless, we can establish a relation between the distance to the first
over-density and the size of the first over-density for leading and
trailing arms separately, which could be observationally
investigated.  A simple linear fit to the data from our simulations
gives the following relations:
\begin{eqnarray}
  \label{eq:combine}
  S_{\rm L} & = & 0.15 \pm 0.01 \times D + 0.2 \pm 0.2, \\
  S_{\rm T} & = & 0.31 \pm 0.04 \times D - 1.7 \pm 0.6.
\end{eqnarray}
Here, S$_{\rm L}$ is the size of the first over-density in leading arm
and S$_{\rm T}$ is the size of the first over-density in trailing arm
and $D$ is the distance (in kpc) to the over-density from the centre of the
satellite. 

\section{Discussion and Conclusions}
\label{sec:disc}

From simulations of dwarf galaxies orbiting a spherical logarithmic and
a NFW dark matter halo (instead of a point mass potential) we find that
the theory of \citet{Kupper2008}, giving theoretical estimates for the
time of formation and the distance to the first over-densities in tidal
tails is over-predicting the results we are obtaining with our simulations.
In case of the location of the first over-density this can be up to factor
of six.  We have shown that this is not due to the fact of the different
sized objects the two studies use but that \citet{Kupper2008} were right
with their estimates as they also used a point-mass potential in their
simulations.  The solution is that one cannot use the simple equations
from their study but has to go back and solve the original epicyclic
equations for the potential one actually uses.  We still see some
discrepancies but those could be explained by the real setup in contrast
to an idealised, approximated theoretical frame-work.  In that respect
we verify the results of \citet{Kupper2008}.

In our study we have shown that the growth-rate of tidal tails (on
circular orbits) does not depend on the potential strength of the
galactic potential as the tails grow with the same rate independent
of the potential used (logarithmic halo or NFW halo) and distance
to the centre of the galaxy.  The tails grow faster if the satellite is
more massive.  We conclude that this is because the simple zero velocity
approximation for the epicyclic equations is in fact not completely valid.
In reality in larger, more massive objects stars escape with higher
relative velocities.

This can be seen as well, when we investigate the growth-rate as function
of the eccentricity of the orbit.  The further in an object orbits, the
stronger is the decrease of the tidal radius at peri-centre.  Therefore,
more stars are able to leave the satellite and escape into the tails.
Naturally, these stars have higher relative velocities and therefore we
grow tidal tails faster when we orbit with higher eccentricity.

Even though we could calculate the rates of growth for both tails,
using eccentric orbits, we do not have sufficient data of the
over-densities, because the code of \citet{Vejar2013} does not consider
the possibility of having more than 2 tidal tails, a feature which
occurs naturally if new stars are lost at peri-galacticon.  The new and
old tails will align with each other when the satellite is approaching
apo-galacticon.

We find a relation between the distance to the first over-density in
the leading and trailing arms and the radius of the orbit and/or the mass
of the satellite.  As the rate of growth does not depend on the radius
of the orbit we conclude that the relation between the distance of the
first over-density and the radius of the orbit must be due to the
change in the tidal radius alone.  This is verified when only changing
the mass of the satellite at the same distance to the galaxy.  A change
in the epicyclic frequency $\kappa$ and the angular velocity $\Omega$
in affecting the results cannot be excluded but are of secondary order.

A very particular result of our study is the visible asymmetry between
the leading and the trailing arm, with the trailing arm growing faster
and being larger than the leading arm.  The most obvious explanation
would be the difference in distance between the two Lagrangian points
in large objects as used in our study.  This is of course an effect
which plays a r\^{o}le but it cannot explain why this trend is more
pronounced at larger radii from the galaxy, where this difference should
become more insignificant.

Furthermore, we find a relation between the size of the first
over-density and the radius of the orbit or the mass of the satellite.
This relation is very clear for trailing arms but not as clear for
leading arms, where the size of the first over-density is very close
to $2$~kpc in all simulations.  A possible explanation for this
behaviour is that particles not always escape with the same velocity
from the satellite but around a central value $v_{\rm esc}$ and with a
small velocity dispersion $\sigma$.  According to the theory
of \citet{Kupper2008} this spread in the escape velocity will lead
to a spread in the value of $y_{\rm c}$ at which the particles turn
around.  Therefore, we will have an extended over-density with a size
$S$.  We expect that satellites with higher masses produce larger
$\sigma$ values and therefore larger over-densities.  Why this is so
strongly visible in trailing arms only is a study on its own and should
be dealt with in a future investigation.

The dependency with the radius is similar. The sizes of over-densities
in the trailing arms grows about three times faster than the respective
size in the leading arm.  Again this obvious asymmetry is not easy to
explain and deserves further studies.

Finally, we combine the results of distance and size of the
over-densities and find a single linear relation between the distance
to the first over-density and its size for leading and trailing arms
separately but independent of the mass of the satellite, the radius of
the orbit and its eccentricity.  This relation could be investigated
with detailed observations of tidal tails. \\

{\bf Acknowledgments:}
MF acknowledges financial support of Fondecyt grant No.~1130521,
Conicyt PII20150171 and BASAL PFB-06/2007.  BR acknowledges funding
through Fondecyt grant No.~1161247.  BR thanks A. Alarcon Jara and D.R.
Matus Carrillo for their help with the code SUPERBOX 
and useful discussions during the realization of this work. 

\bibliographystyle{mn2e}

\label{lastpage}

\end{document}